\documentclass[11pt]{article}
\usepackage{graphicx}

\newsavebox{\PSLASH}
\sbox{\PSLASH}{$p$\hspace{-1.8mm}/}

\begin{document}
\title{Scaling Relations for Contour Lines of Rough Surfaces  }
\author{M. A. Rajabpour$^1$\footnote{e-mail: rajabpour@ipm.ir} ~and S. M. Vaez Allaei$^2$\footnote{e-mail: smvaez@ut.ac.ir} \\ \\
$^1$ Institute for Studies in Theoretical Physics and Mathematics,
Tehran 19395-5531, Iran\\
$^2$ Department of Physics, University of Tehran, Tehran 14395-547, Iran\\
} \maketitle
\begin{abstract}
Equilibrium and non-equilibrium growth phenomena, e.g., surface growth,
generically yields self-affine distributions. Analysis of statistical
properties of these distributions appears essential in understanding
statistical mechanics of underlying phenomena. Here, we analyze scaling
properties of the cumulative distribution of iso-height loops (i.e.,
contour lines) of rough self-affine surfaces in terms of loop area and
system size. Inspired by the Coulomb gas methods, we find the generating
function of the area of the loops. Interestingly, we find that,  after
sorting loops with respect to their perimeters,
Zipf-like scaling relations hold for ranked loops. Numerical simulations are also provided in order to demonstrate the proposed scaling relations.

\vspace{5mm}
\noindent PACS number(s): 47.27.eb,
\end{abstract}
\section{Introduction}\

Self-affine distributions are ubiquitous in many phenomena in nature,
such as in growing surfaces and interfaces
\cite{stanley,Kondev1,Iraji,Isichenko}, fractured media \cite{Sahimi,drazer},
and graphs of two-dimensional turbulent flows
\cite{Isichenko,ramshan}.
Self-affine distributions have also been used as a tool to
study scaling properties of two-dimensional statistical models by
mapping these models to a two-dimensional Coulomb gas \cite{nienhuis, kondev3}. Moreover, crystal growth, the growth of bacterial colonies,
and the formation of clouds in the upper atmosphere \cite{Pel} are
all examples of non-equilibrium phenomena which grow self-affine
rough surfaces. The above applications on a fundamental level make
the surface-growth problem as a paradigm for a broad class of problems in
the context of non-equilibrium statistical mechanics.

Self-affine surfaces can be described by their height distribution
function. From statistical point of view, it is necessary to explore
topography of this kind of surfaces. In such surfaces, heights are
invariant under re-scaling, namely $h(\textbf{r})\cong
b^{-H}h(b\textbf{r})$, where $H$ is called the roughness exponent or
the
\textit{Hurst} exponent. It implies that in a self-affine surface,
the variance of the surface height, i.e.,
$\sqrt{\langle[h(\textbf{x})-\langle{h}\rangle]^2\rangle}$, scales
as $L^{H}$, where $L$ is the size of the system and average is taken
over $x$. If we require translational and rotational
invariance of the surface then the structure function of this
surface behaves as
\begin{eqnarray}\label{structure function}
C_{2}(\textbf{r})=\langle[h(\textbf{x})-h(\textbf{x}+\textbf{r})]^{2}\rangle\sim
|\textbf{r}|^{2H}.
\end{eqnarray}
The above equation gives a simple formula to calculate the
roughness exponent. To determine that a given surface is self-affine
or multi-affine we need to measure the $p$th order structure
function defined by
$C_{p}(\textbf{r})=\langle|h(\textbf{x})-h(\textbf{x}+\textbf{r})|^{p}\rangle$.
The exponent hierarchy $\alpha_p$ is defined through the relation
$C_{p}(r)\simeq r^{\alpha_{p}H}$. The exponent $\alpha_p$ varies
linearly with $p$ for a self-affine surface. For a multi-affine
surface, instead, it would vary non-linearly with $p$ \cite{Barabasi}. The
Fourier space counterpart of the structure function is Fourier power
spectrum $S(\textbf{q})=\langle|h(\textbf{q})|^{2}\rangle$, where
$h(\textbf{q})$ is the Fourier transform of $h(\textbf{r})$.
Equation (\ref{structure function}) gives
the scaling relation for the power spectrum, i.e.,
$S(\textbf{q})\sim
|\textbf{q}|^{-2(1+H)}$, for small values of $q$ or large values of
$r$. One way to generate a Gaussian ensemble of self-affine surfaces
is by taking each Fourier height as an independent Gaussian random
variable with variance given by $S(\textbf{q})\sim
|\textbf{q}|^{-2(1+H)}$. In other words,
\begin{eqnarray}\label{free energy}
P\{h\}\sim \exp\left[-\frac{k}{2}\int_{0}^{\Lambda}d^{2}\textbf{q}\
\textbf{q}^{2(1+H)}h_{\textbf{q}}h_{-\textbf{q}}\right],
\end{eqnarray}
where $\Lambda=1/a$ is the high-momentum
cutoff and $k$ is the
stiffness. A family of self-affine surfaces having all the required
properties can be generated by the above distribution. For rough
surfaces with unbounded heights we have $0\leq H<1$,
where the higher $H$ is related to  smoother surface with hills.
In a self-affine distribution $H>1/2$ ($H<1/2$),
it implies positive (negative) correlations among the increments of the
generated values,
$H=1/2$ means that the statistics of the surface follows that of
a Brownian motion. At $H=0$,
it is possible to write
Eq.~(\ref{free energy}) in the real space by using ordinary derivative
$P\{h\}\sim \exp\left[-\frac{k}{2}\int_{0}^{L}d^{2}\textbf{x}
(\partial_{\textbf{x}} h)^{2}\right]$. For the general case we
should replace ordinary derivative with the fractional one, that is,
$P\{h\}\sim \exp\left[-\frac{k}{2}\int_{0}^{L}d^{2}\textbf{x}
h(-\nabla^2)^{1+H} h\right]$, where the fractional derivative is
defined by $(-\nabla^2)^{1+H}e^{i{\bf q}.{\bf
x}}=-|q|^{2+2H}e^{i{\bf q}.{\bf x}}$ (for more details see
\cite{Herman}).

The contour lines that are generated by a cut through the surface at
a certain height are important in characterizing
self-affine surfaces.

In Fig.~$1$ we plotted an example of the set of contour lines of a
rough surface. The statistical properties of contour lines of rough
surfaces show fractal behavior.
The accepted fractal dimension of a contour line $D=\frac{3-H}{2}$
was found by Kondev and Henley (KH)
by using scaling arguments \cite{kondev2}. Recently,
Schramm and Sheffield \cite{Schramm} proved rigorously that the contour
lines of Gaussian free field with $H=0$ are conformally invariant
with fractal dimension $D=\frac{3}{2}$, which is in agreement with
the KH result.
Conformally invariant curves in statistical
physics can be investigated by Coulomb gas field theory
\cite{nienhuis}. The most well-known loop model that can be
investigated by this field theory is the $O(n)$ model,
which can be
defined on the honeycomb lattice as follows: take the ensemble of
loops on the honeycomb lattice so that the generating function of the
model is given by $Z=\sum x^{l}n^{N}$, where $N$ and $l$ are 
number of the
loops and bonds, respectively, $n$ is the weight of each
loop, and $x$ is the weight of each bond.
At the critical point, this loop model can be
investigated, after mapping the loop model to
the solid on solid (SOS)
model \cite{nienhuis}, by the free field theory
$P\{h\}\sim \exp[-\frac{k}{2}\int_{0}^{L}d^{2}x (\partial h)^{2}]$.
It is also possible to find the scaling exponents of conformal
curves by the above field theory \cite{nienhuis}.
\begin{figure}
\begin{center}
\includegraphics[angle=0,scale=0.8]{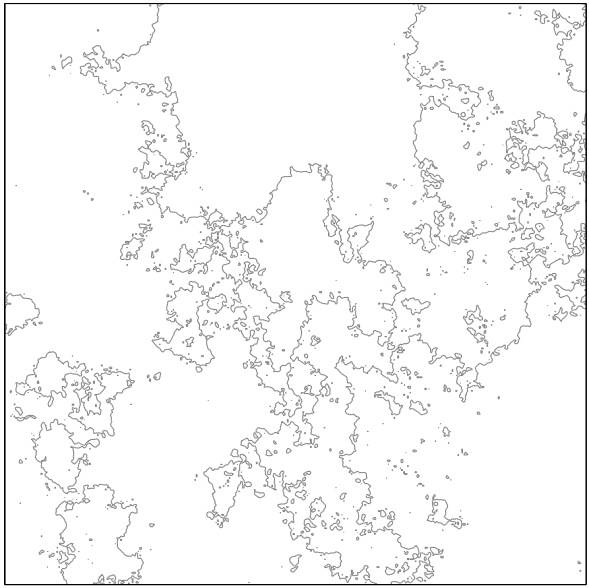}
\caption{Small part of contour lines of a rough surface with size
$3000^{2}$ and $H=0.5$; by zooming in on the picture one can see
many small loops.}
 \end{center}
\end{figure}
Since the height ensemble of a rough surface is not conformally
invariant, rigorous investigating of their contour lines is more
difficult than the Coulomb gas case. Indeed, one can not employ the
powerful tools of conformal field theory (CFT) to study this system.
For a rough surface with a generic $H$ there is no rigorous proof
for results obtained by KH \cite{Schwartz}. Nonetheless, it seems
that the contour line ensemble shows scaling properties similar to
the conformal curves encountered in some models such as the contour
lines of tungsten oxide ($WO_3$) \cite{Rajabpour} and KPZ surfaces
\cite{saberi}.

In this paper, by using
techniques which are common in the realm of Coulomb gas field
theory, we introduce new scaling laws for some properties of contour
lines of self-affine rough surfaces. The scaling properties of the
cumulative distribution of the number of contours versus the area of
the contours and the size of the system
are also obtained. In addition,
 we find a close relation between the
cumulants of $A$, the area of contour lines, and the eigenvalues of
the fractional Laplacian. Finally,
we introduce the scaling property of
ranked contour lines versus both rank and system size
(Zipf's law). Numerical simulations are also provided to substantiate our analysis.

\section{Numerical methods}\

To generate self-affine rough surfaces in our numerical simulations,

we have used the successive
random addition method \cite{Voss}. In our simulations we have
generated surfaces of size $L\times L$ with
$L\in\{400, 600, 800, 1200, 2000, 3000, 4000\}$.
To investigate the effect of roughness
exponent on the scaling relations we used several values of
$H\in\{0.3, 0.4, 0.5, 0.6, 0.7\}$.
In each case, all calculations have been averaged
over 200 realizations.

To generate the loops in the contour lines we used
a contouring algorithm that treats the input matrix as a regularly spaced grid.
The algorithm scans this matrix and compares the values of each block
of four neighboring elements (i.e., a plaquette) in the matrix to
the contour level values. If a contour level falls within a cell,
the algorithm performs a linear interpolation to locate the point at
which the contour crosses the edges of the cell. The algorithm
connects these points to produce a segment of a contour line. After
generating the contours of a given surface,
in order to eliminate the effect of the edges of the lattice we have
excluded the contours crossing the edges of the lattice.
\begin{figure}
\begin{center}
\includegraphics[angle=0,scale=0.6]{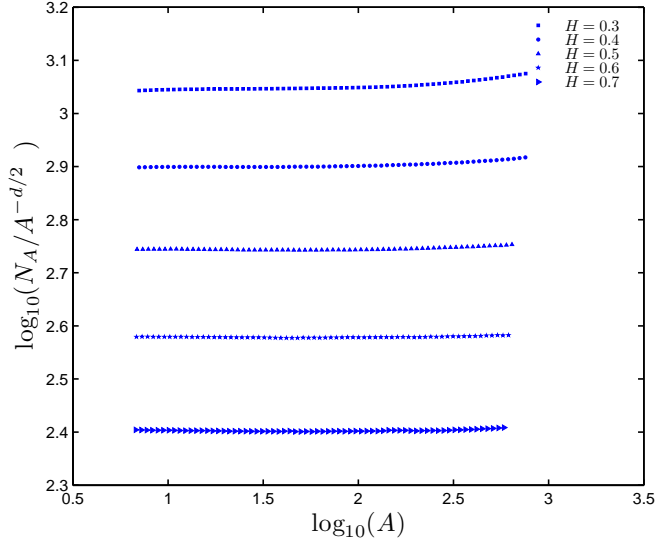}
\caption{Scaling relation for cumulative distribution of areas.
Curves show $\frac{N_A}{A^{-d/2}}$ for rough surfaces with size
$4000^{2}$ and different roughness exponents.}
 \end{center}
\end{figure}
To show the goodness of the fits and consistency of our simulations
with theory, we used the following three different methods for
estimating the exponents: (\textit{a}) we numerically calculated
local slops of the curves by fourth-order numerical differentiation
for non-uniform data points; e.g., in the case of
Eq.~(\ref{cumulative}), derivation of $\log_{10}(N_{A})$ relative to
$\log_{10} A$. (\textit{b}) We present some of the curves by
dividing both sides of a scaling relation to the claimed power law
to show how seriously they are aligned or how they deviate from a
horizontal line, e.g., Fig. 2. And, (\textit{c}) We used Bayesian
analysis without prior distribution, namely Likelihood analysis
\cite{movahed1,movahed2,movahed3,co04} to calculate the accuracy of
the exponent generated from our numerical results.

\section{Cumulative distribution of area}\
\setcounter{equation}{0} A key difference between the contour lines
in Coulomb gas field theory and the self-affine rough surfaces is in
the fractal dimension of the set of all contour lines. For a given
self-affine rough surface, this fractal dimension is $d=2-H$. It is
well-known that many of the scaling relations in Coulomb
gas field theory remain unchanged just by substituting this $d$ as
the dimension of our set.
To give an example, let us define the fractal
dimension of a contour line $D$ as
$\textit{l}\sim\textit{R}^{D}$, where $\textit{l}$ is the the
perimeter of the
contour and $\textit{R}$ is the radius of gyration.
Moreover, the probability of finding a contour loop with length
$\textit{l}$ is $N_{\textit{l}}\sim\textit{l}^{-\tau}$. One can show
that there is a hyperscaling relation between the scaling exponents
$D$ and $\tau$ as follows:
\begin{eqnarray}\label{hyperscaling}
D(\tau-1)=d,
\end{eqnarray}
which is exactly the same as the hyperscaling relation for domain
walls in statistical models \cite{Cardy1}. Following KH, the cumulative
distribution of the number of contours with area greater than $A$
 has the following form:
\begin{eqnarray}\label{cumulative}
N_{A}\sim \frac{C}{A^{{d}/{2}}}.
\end{eqnarray}
This gives the right answer for Coulomb gas loops with zero
roughness exponent \cite{Cardy1}. In the rest of the paper using new
conjectures we will demonstrate some other evidences to support the
above relation. This in turn leads us to several new scaling
relations.

We checked Eq.~(\ref{cumulative}) by using numerical simulations
for different $H$'s,
see Fig $2$. As  is shown, we plot $\log_{10}(\frac{N_{A}}{A^{-d/2}})$
versus $\log_{10}(A)$ to show how seriously they follow
Eq.~(\ref{cumulative}).
The straight horizontal curves exhibit that the proposed scaling
relation is preserved up to $2$ orders of magnitude of $A$.
As is seen, in the case of $H=0.3$ we have a
small deviation from the proposed exponent at large values of $A$,
which is related to finite-size effects.
For a given lattice
size and for small values of $H$, there are not so many large contour
lines, but we have many small ones. This is led
by the nature of self-affinity at small \textit{Hurst} exponents.
There are no deviations when we increase $H$ (Fig.~2).
In Table~\ref{tab1}, we report the best fit
values calculated by the likelihood analysis
\cite{movahed1,movahed2,movahed3,co04} at $68.3\%$ and $95.4\%$
confidence levels.
\newline
\begin{table}[htp]
\begin{center}
\begin{tabular}{|c|c|c|c|}\hline
$H$& Theory &  Local exponent ($1\sigma$)  & Local exponent ($2\sigma$) \\
\hline
$0.3$ & -0.850 &$-0.840 \pm {0.010}$ &$-0.840 \pm {0.020}$    \\
\hline
$0.4$ & -0.80 &$-0.795 \pm {0.006}$ & $-0.795 \pm {0.020}$
  \\ \hline
$0.5$ & -0.750 &$-0.752 \pm {0.010}$ & $-0.752 \pm {0.025}$
\\ \hline
$0.6$ & -0.700 &$-0.703 \pm {0.005}$ & $-0.703 \pm {0.020}$
\\ \hline
$0.7$ & -0.650 &$-0.652 \pm {0.005}$ & $-0.652 \pm {0.020}$\\
\hline
\end{tabular}
\end{center}
\caption{\label{tab1} The best fit values of 
exponent $d/2$
derived by using the likelihood method. $\sigma$ denotes
standard deviation of each calculated exponent.}
\end{table}

For loops
corresponding to surfaces with $H=0$, using 
Coulomb gas techniques, Cardy and Ziff showed that $C$ has the
universal form as a function
of the system size $L$ for different critical statistical physics
models \cite{Cardy2}. To calculate $C$, Cardy and Ziff
evaluated the total area inside all loops using two different
methods, and then they found the universal form of $C$. Inspired by this
method, we argue to give some new scaling relations for contour
lines of self-affine rough surfaces.
Using Eq.~(\ref{cumulative}) it is straightforward
to show that $\langle A_{tot}\rangle=CL^{H}$,
for $0<H<1$, and for $H=0$ it has a logarithmic form.

Let us consider a typical point $x$ with height $h$ above
the horizon (we cut our self-affine surface by a plane). If we draw a
circle of radius $h^{{1}/{H}}$ around the point,
since we are dealing with a rough surface, all points inside
the circle will be above the horizon.
In other words, inside the loop is a compact region with the fractal
dimension 2. Since the fractal dimension of the clusters is 2,
one could obtain the total area of the clusters
proportional to the area of the system.
This is just a lower bound
for the interior areas of the loops ---
see \cite{Olami}. In addition,
it is also possible to see from simulation that by cutting
the surface from the average height one could get always clusters of
the order of the system size.
Thus one can get the following scaling
relation for $C$ with respect to the system size:
\begin{eqnarray}\label{C}
C\sim L^{2-H}.
\end{eqnarray}
This indicates that the number of contour lines with area greater than
$A$ per total length of all contours, i.e., $L^{2-H}$, is independent
of the system size. It is also worth noting
that the cumulative distribution of the contours with area $A$
is independent of $a$, the ultraviolet
cutoff, which is another length scale. Our simulations confirm the
validity of the scaling
relation (\ref{C}) for different values of $H$, see Fig.~$3$.

The above result is also useful to get another nontrivial equation
for contour lines. To calculate the total area we can use the
formula
\begin{eqnarray}\label{Atot}
\langle A_{tot}\rangle=\int<d(r)>\textit{d}^{2}\textit{r},
\end{eqnarray}
in which $d(r)$ is the minimum number of loops which must be crossed
to connect $r$ to the edge of the lattice. Since the total area
inside the loops is proportional to the area of the system,
we conclude
\begin{eqnarray}\label{d(r)}
d(r)\sim(\frac{r}{L})^{2H}.
\end{eqnarray}
This is reminiscent of the height correlation function in the
self-affine rough surfaces. For $H=0$ the relation is logarithmic
and was proved explicitly in \cite{Cardy2}.
\begin{figure}
\begin{center}
\includegraphics[angle=0,scale=0.6]{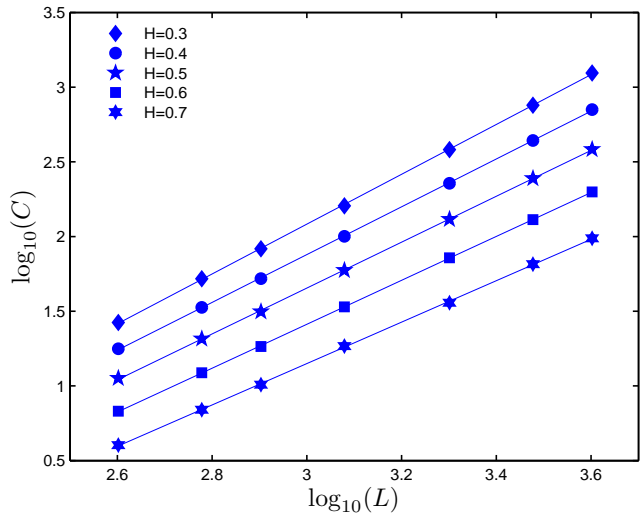}
\caption{Scaling relation for the coefficient of cumulative
distribution of areas for rough surfaces with respect to system size
for different values of $H$ as shown in the graph. Slops of the
curves from top to bottom are given by $1.66\pm 0.04$, $1.60\pm
0.03$, $1.53\pm 0.03$, $1.46\pm 0.04$, and $1.37 \pm0.04$.}
 \end{center}
\end{figure}
These results show that one may investigate contour loops of rough
surfaces by defining currents for the loops. Again,
by analogy with the Coulomb gas methods one can define
$\textit{J}_{\mu}(x,y)\sim\sqrt{g}\frac{\epsilon_{\mu\nu}\partial_{\nu}h}{L^{H}}$
as the current density of loops. This is a natural candidate if we
imagine that the height function is extended to the
two-dimensional manifold in such a way that it is constant within each plaquette.
Normalization with respect to width is necessary because we have a
rough surface where width is changing by size. This definition for
the current density means that we can map our height model to the
contour lines or vice versa. Since iso-height lines have the same role as the domain
walls between the positive and negative heights, the directional
derivative of $h$ along a contour must be zero and it must vary
along a line normal to the contour. Using the above function to
parameterize the geometry of contour line, it is possible to write
\begin{eqnarray}\label{current}
A=-\frac{1}{2}\int\int|x-x'|\delta(y-y')J_{y}(x,y)J_{y}(x',y')d\textbf{r}d\textbf{r}'.
\end{eqnarray}
This equation is independent of our 
definition of currents.
By using simple dimensional analysis, it is
not difficult to find our special normalization, i.e.,
 $\frac{1}{L^{H}}$. One can check that
 Eq.~(\ref{current}) gives $<A_{tot}> \sim L^{2}$. Using
Eq.~(\ref{current}), inspired by Cardy's argument \cite{Cardy1}, we
find the generating function of the cumulants of area of contour
loops. The argument for getting cumulants of area is as follows.
For the simplicity, we use the
Dirichlet boundary condition, $h=$const, on the boundary of the
system, which means that loops do not cross the boundary.
After integration by parts,
Eq.~(\ref{current}) gives $A\simeq \int d^{2}r h^{2}(r)$. In simulation and experiment there are many
curves emerging from the boundary and going back to another point in
the boundary; therefore, there will be no exact Dirichlet boundary
condition.
\begin{figure}
\begin{center}
\includegraphics[angle=0,scale=0.6]{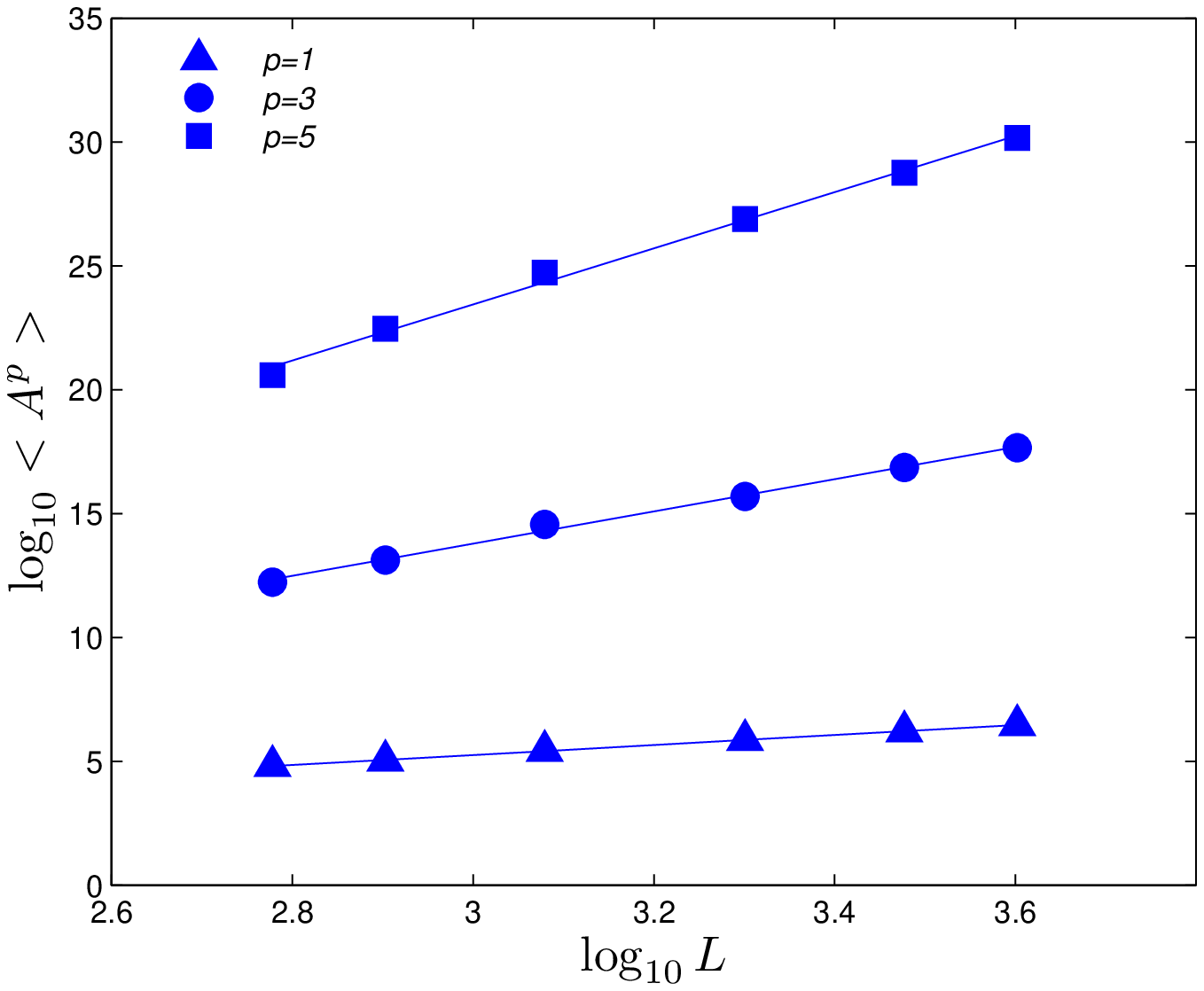}
\includegraphics[angle=0,scale=0.6]{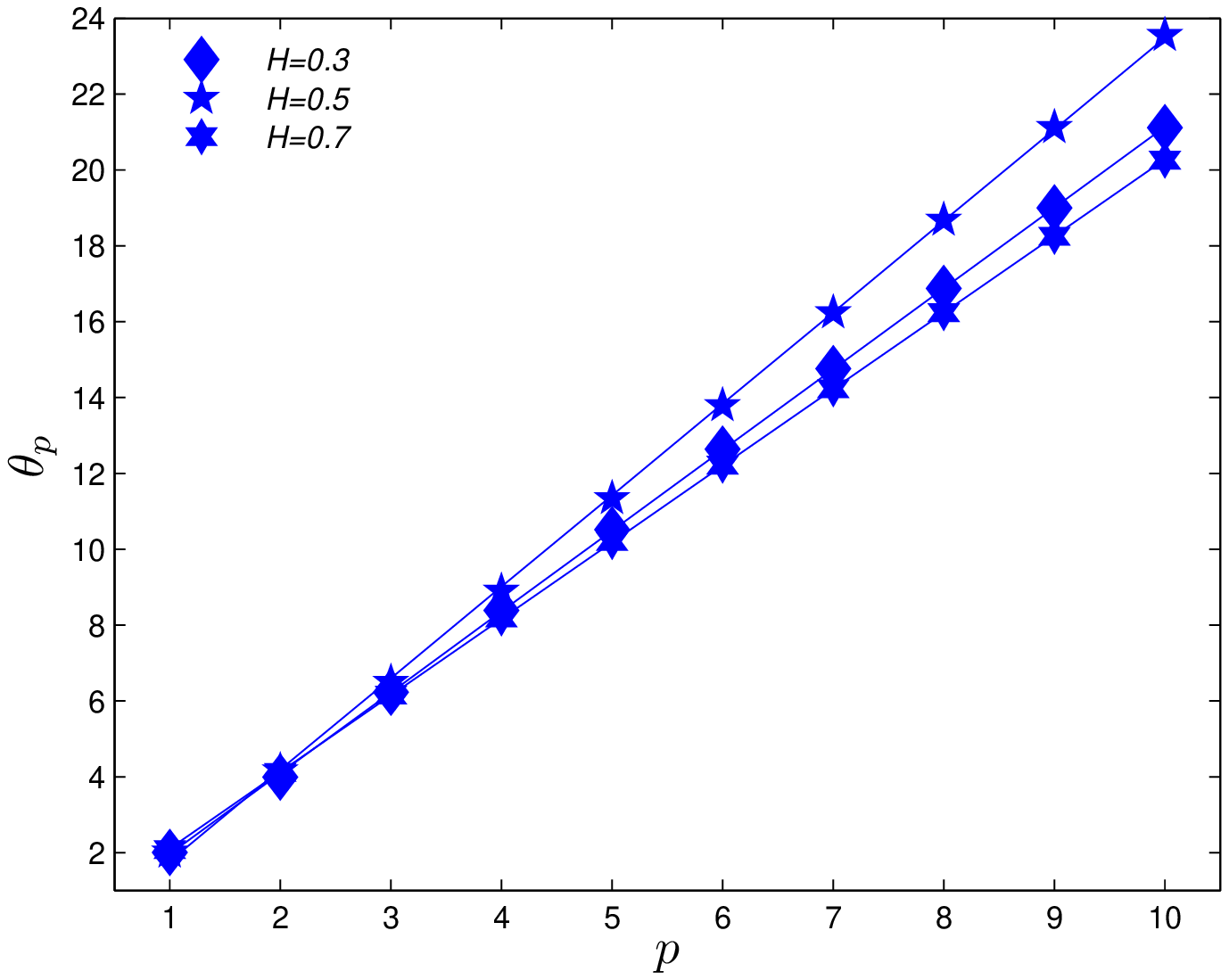}
\caption{Top: Moments $\left\langle A^p \right\rangle$ of loop areas
of rough surfaces versus system size; for $p=1, 3$, and $5$. Here we
have the exponents $\theta_{1}=2.02\pm 0.06$, $\theta_{3}=6.05\pm
0.55$, and $\theta_{5}=11.03\pm 1.06$.\hspace{.2cm} Bottom:
$\theta_{p}$ versus $p$ for surfaces with size $4000^{2}$ with
different roughness exponents $H$ as indicated on the graph.}
 \end{center}
\end{figure}

However, as we will see in the simulations, many of our scaling
relations, especially the distribution of contours, are independent of
the boundary conditions. By using the real space representation of the
height distribution and the Gaussian integral,
one can derive
\begin{eqnarray}\label{generating function}
\langle
e^{-uA}\rangle=\frac{\det(-\partial^{2+2H}+\frac{2gu}{kL^{2H}})^{-1/2}}{\det(-\partial^{2+2H})^{-1/2}}.
\end{eqnarray}
where $k$ is the stiffness and $u$ is an auxiliary field.
One can write the above equation as an infinite
sum by using Fourier transform
\begin{eqnarray}\label{generating function11}
\langle e^{-uA}\rangle= e^{-\frac{1}{2}\sum_{m}
\ln\left(1+\frac{2guL^{2}}{k\lambda_{m}}\right)},
\end{eqnarray}
where $\frac{\lambda_{m}}{L^{2+2H}}$ are the eigenvalues of the
fractional Laplacian with Dirichlet boundary conditions
\cite{Herman}. Expanding 
Eq.~(\ref{generating function11})
gives the higher cumulants of $A$,
\begin{eqnarray}\label{moment}
\langle A^{p}\rangle \sim L^{2p}\sum_{m}\frac{1}{\lambda_{m}^{p}}.
\end{eqnarray}
The sum is convergent for all values of $p$ and $H$ except
$p-1=H=0$, which is logarithmic with respect to $L$.
To check the above equation we calculated $\langle A^{p}\rangle$
for surfaces with different roughness exponents and different sizes.
For $p=1$ all of the surfaces have $\langle A^{1}\rangle\sim
L^{\theta}$ with $\theta=2\pm 0.05$. For higher moments one can
write $\langle A^{p}\rangle \sim L^{\theta_{p}}$ with
$\theta_{p}\sim \theta_1 p$. For surfaces with roughness exponent
between $0.3$ and $0.7$,
the exponent $\theta_{1}$ varies
from $1.94$ to $2.15$ . One can see in Fig.~4b that all of the
$\theta_{p}$'s are linear with respect to $p$. The deviation from
$\theta=2$ could be related to our restriction in getting larger
sizes in simulation.

\section{Zipf's law for contour lines}\
\setcounter{equation}{0}

Another interesting scaling relation is the universality of the
distribution of the ranked loop perimeters,
which is named Zipf's law
\cite{Mandelbrot}. Following \cite{Mandelbrot,jan},
the average perimeter of the $\textit{n}$th largest cluster can be found by
Eq.~(\ref{cumulative}),
which is called by Mandelbrot the Zipf distribution
\begin{eqnarray}\label{rank perimeter}
\textit{l}_{\textit{n}}\sim \frac{L^{D}}{n^{\frac{D}{d}}},
\end{eqnarray}
where $d$ is the fractal dimension of all loops and $D$ is the
fractal dimension of one of loops.
We should emphasize that we have normalized the equation
with the appropriate power of total number
of contour loops, so we ignore here the scaling of
the total number of loops \cite{Mandelbrot}.
We have numerically checked this scaling
relation for self-affine surfaces,
both with respect to rank $n$ and the system size $L$. As shown in
Fig.~$5$, in three subfigures (for $H\in\{0.3, 0.5, 0.7\}$) for
eight different realizations, we presented
the log-log plot of $\frac{l_n}{n^{-D/d}}$ versus
$n$. Here, ${l_n}$ shows a scaling relation according to
Eq.~(\ref{rank perimeter}). For the case of $H=0.3$, the scaling
relation is preserved for over 2 orders of magnitude of $n$.
Since the number of small loops is few, in larger values of $H=0.5,0.7$,
we could see the agreement just for 1 order of magnitude.
\begin{figure}
\begin{center}
\includegraphics[angle=0,scale=0.4]{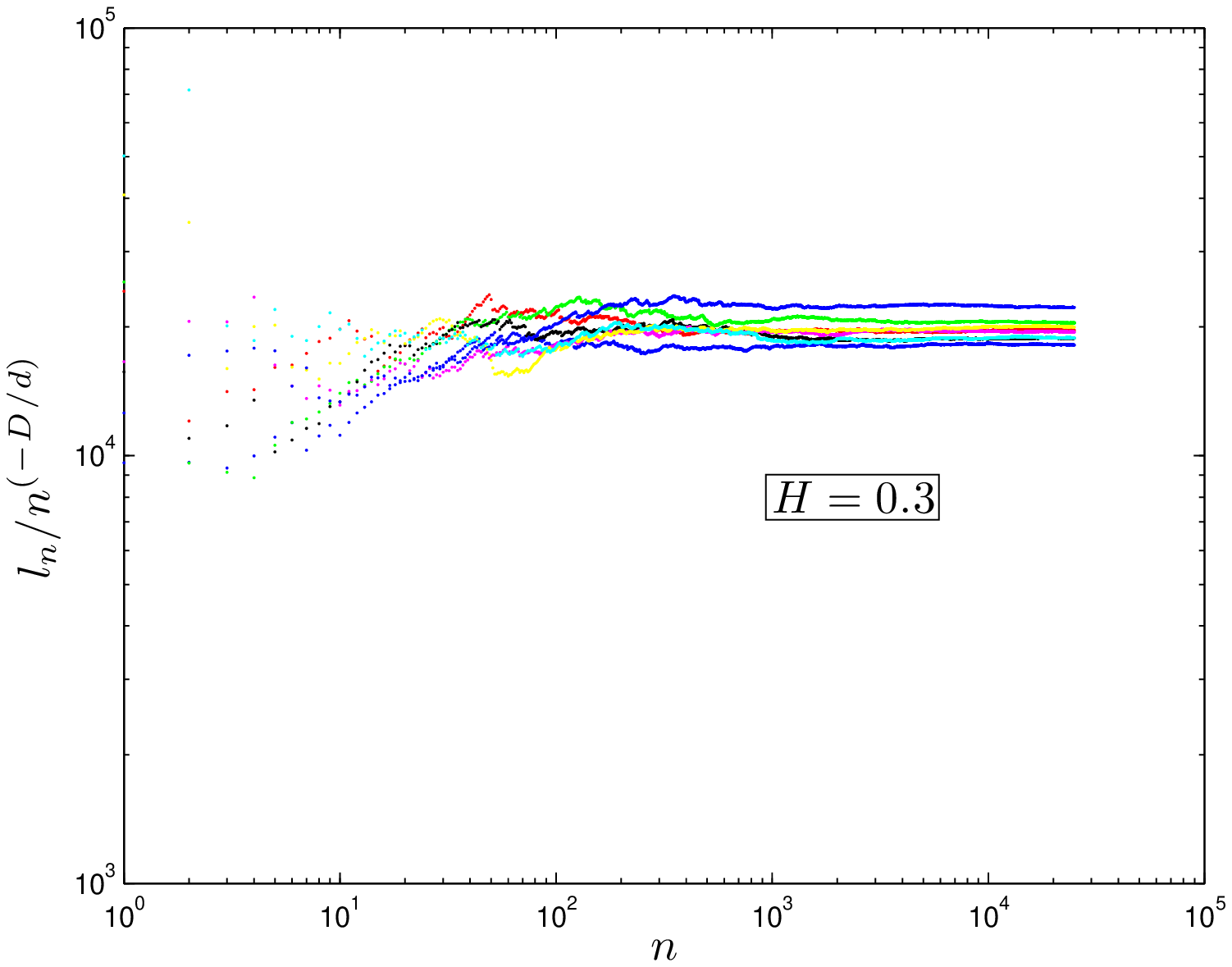}
\includegraphics[angle=0,scale=0.4]{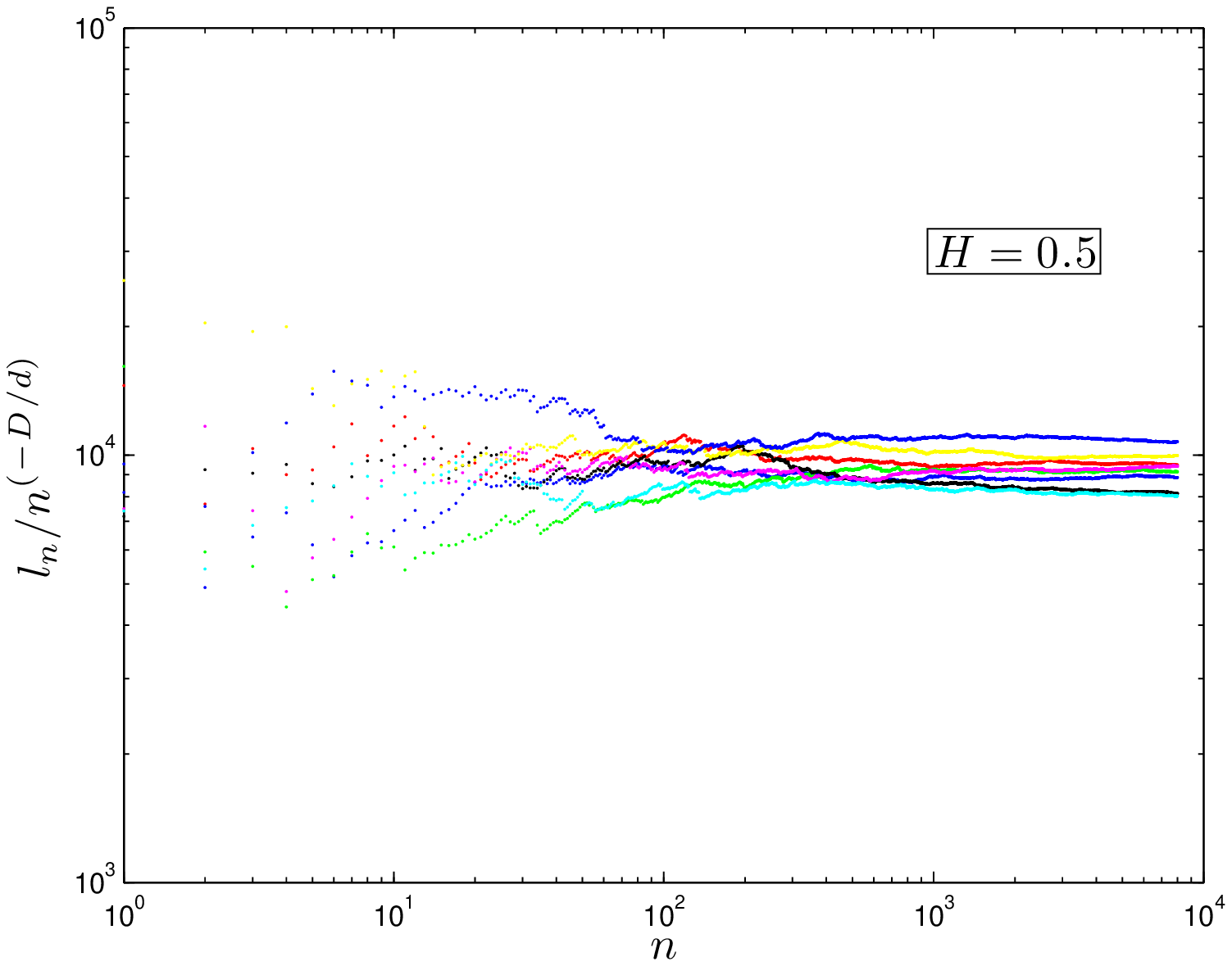}
\includegraphics[angle=0,scale=0.4]{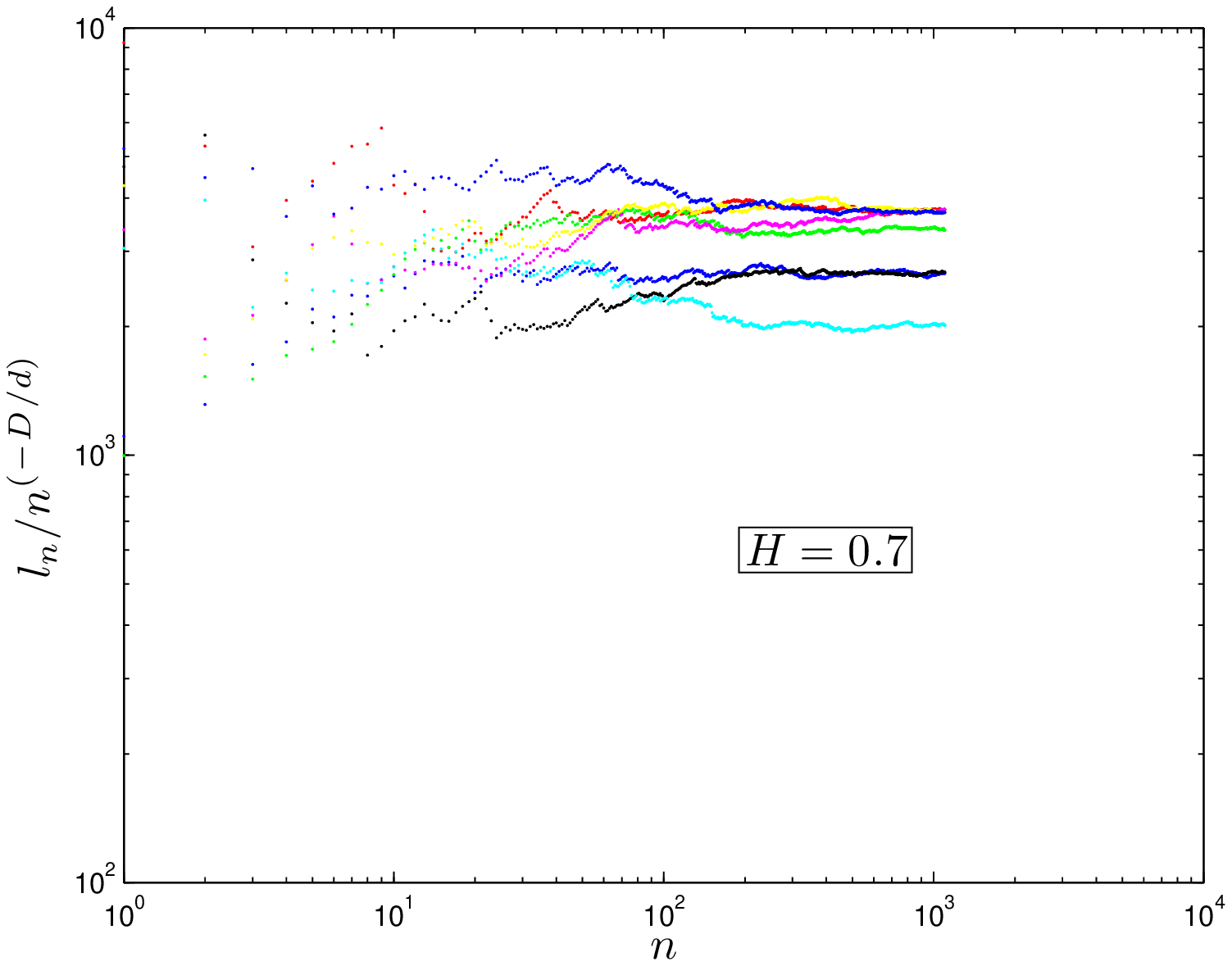}
\includegraphics[angle=0,scale=0.4]{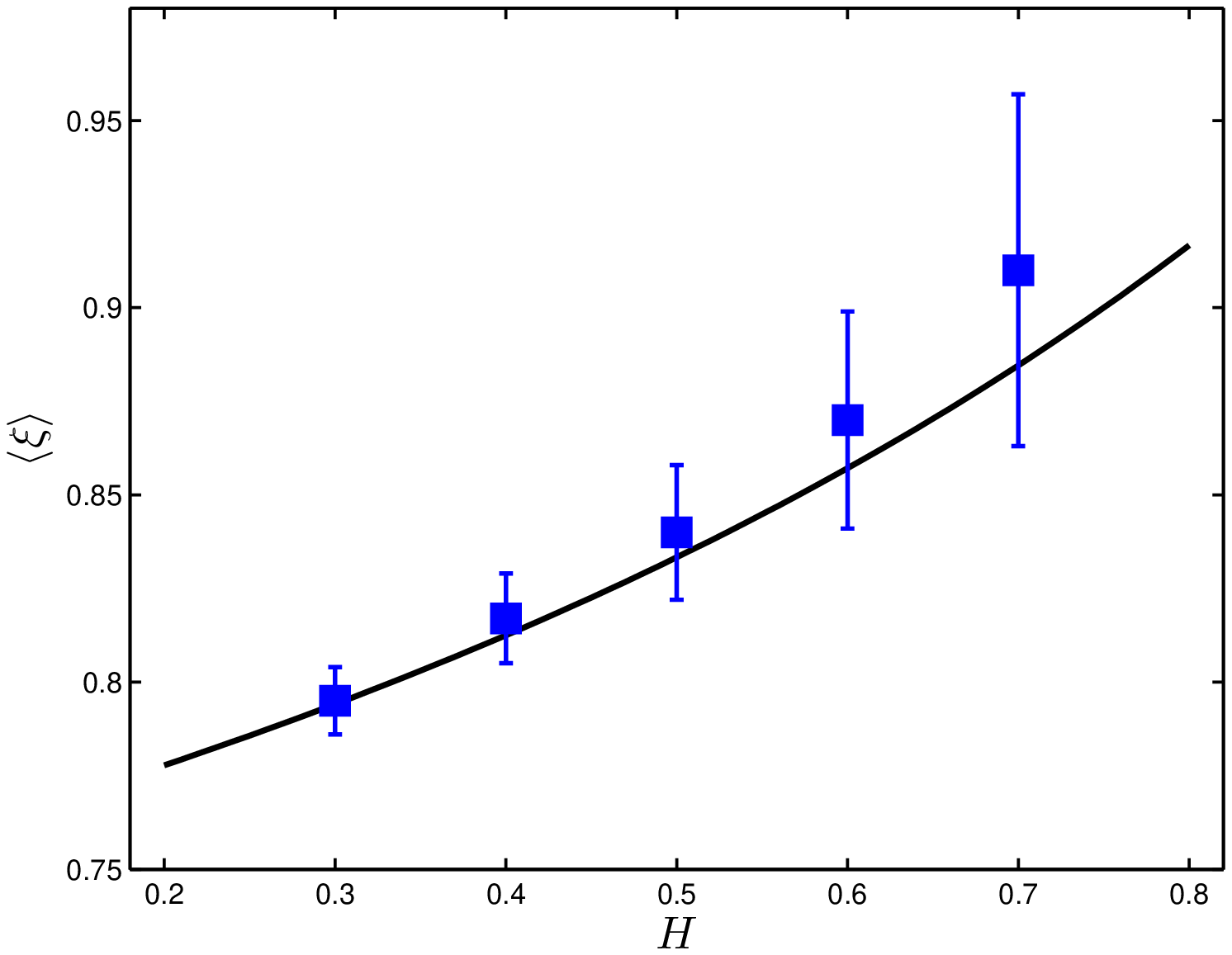}
\caption{Top left ($H=0.3$), right($H=0.5$), and bottom left
($H=0.3$): for system size of $L=4000$, the curves of ranked loop
perimeters divided by $n^{-D/d}$ vs rank numbers are shown for eight
different realizations. Bottom right: the squares stand for the
averaged $\xi$ (the numerical estimated exponent) over $200$
realizations. The solid line shows the theoretical relation between
$\xi$ and $H$, i.e., $\xi=\frac{D}{d}$.}
\end{center}
\end{figure}
To calculate the exponent in our numerical results
let us consider $l_n\sim n^{-\xi}$. We calculated the exponent;
Fig.~ $5$ (bottom right) depicts the variation of $\xi$ versus
$H$ (the average is over different realizations). Since in higher $H$s we have
lower number of loops, thus $\xi$ of higher values of $H$ have lower accuracy.
We also numerically checked the
relation of $\textit{l}_{\textit{n}}$ versus
$L$. In the case of $H=0.3$,
we obtained a scaling relation with exponent $1.38\pm0.03$, which is
near theoretical value $1.35$. In higher values of $H$,
the estimated exponents are not sufficiently accurate
because the number of loops in smaller system sizes is low.
With the same method one can find the average area and
the radius of gyration as a function of rank
\begin{eqnarray}\label{rank}
A_{\textit{n}}\sim
\frac{L^{2}}{n^{\frac{2}{d}}},\hspace{2cm}R_{\textit{n}}\sim
\frac{L}{n^{\frac{1}{d}}}.
\end{eqnarray}
Both of the above formulas are in good agreement with our numerical
results.
In these kinds of scaling relations the error of the estimated exponents for
large system sizes are considerably small. We believe Eqs~(\ref{rank perimeter})
and (\ref{rank}) provide a good method to calculate the fractal
dimension of a single contour as well as the fractal dimension of all contours.

\section{Discussion and Conclusion}\
In summary, by using field theory of rough surfaces and considering
current for the model, we confirmed the previously known scaling relation
for cumulative distribution of area. In addition, we found a new scaling
relation for this distribution with respect to system size.
Since the action is not translationally invariant and
the small momenta are important, naturally scaling properties depend
to the system size. It seems that large momenta do not contribute in
the scaling properties. Although system is not invariant under
homogeneous translation, it is not difficult to see that it is invariant under $h\rightarrow h+\epsilon_{\mu\nu}a^{\mu}x^{\nu}$, which means it is
inhomogeneously translational invariant. Using inhomogeneous
translation one can define the currents
$\textit{J}_{\mu}(x,y)\sim\epsilon_{\mu\nu}\partial^{1-H}_{\nu}h$
corresponding to Wilson loops of the theory and re-derive the results
of Sec.  ~II.
Since we only investigated the scaling properties of the contour lines,
these two different given currents lead to the same scaling relations.

Considering these currents for contour lines we think that there may be
a close relation between the statistics of these lines and the
eigenvalues of fractional Laplacian. In this paper, we discussed leading
scaling behavior with respect to the system size, however, to see the effect
of the eigenvalues of the fractional Laplacian one needs more careful study of
the amplitudes as well. Since there is no conformal invariance in the height
ensemble, finding the exact values of $d(r)$ and $C$ using the techniques of
the Coulomb gas is not tractable.

We confirmed our proposed scaling relations by simulations through cutting a
self-affine surface at different heights. We have only interpreted
the results for the case of cutting the surface at its mean height.
But we checked also all of the scaling
relations for the cases of cutting the surface at heights
$h=\{0.1,0.2,0.3,0.4,0.5,0.6,0.7,0.8,0.9\}\sigma$,
where $\sigma$ is the height variance of the surface. We have not
seen any meaningful deviation from what we obtained for the mean
height.

We also introduced new Zipf-like scaling relations for the contour
lines of self-affine rough surfaces, and verified them
via simulations. We believe the same scaling relations are applied
to the clusters of rough surfaces but with different exponents.

\begin{center}
{\bf ACKNOWLEDGMENTS}
\end{center}

The work of S.M.V.A. was supported in part by the Research Council of
the University of Tehran. We thank A. Rezakhani Tayefeh, M. Habibi,
H. Mohseni Sadjadi,
A. Naji, M. Nouri-Zonoz, and M. Yaghoubi for useful discussions.
We are grateful of N. Abedpour, M. F. Miri, and M. Sadegh Movahed
for useful comments. We are also indebted anonymous referees for
their enlightening comments.


\begin{thebibliography}{99}

\bibitem{stanley} A. -L. Barabasi and H. E. Stanley, \textit{Fractal
Concepts in Surface Growth} (Cambridge University Press, New York, 1995).

\bibitem{Kondev1}J. Kondev, C. L. Henley, and D. G. Salinas, Phys.
Rev. E \textbf{61}, 164 (2000).

\bibitem{Iraji} A. I. Zad, G. Kavei, M. R. R. Tabar, and
S. M. V. Allaei, J. Phys: Condens. Matter \textbf{15}, 1889 (2003).

\bibitem{Isichenko} M. B. Isichenko, Rev. Mod. Phys. \textbf{64}, 961 (1992).

\bibitem{Sahimi} M. Sahimi, Phys. Rep. \textbf{306}, 213 (1998).

\bibitem{drazer}  G. Drazer, H. Auradou, J. Koplik, and J. P. Hulin,
Phys. Rev. Lett. \textbf{92}, 014501 (2004).

\bibitem{ramshan} R. Ramshankar and J. P. Gollub, Phys. Fluids A \textbf{3}, 1344 (1991).

\bibitem{nienhuis} B. Nienhuis, in \textit{Phase Transition and Critical
Phenomena}, edited by C. Domb and J. L. Lebowitz (Academic, London, 1987)
, Vol. 11.

\bibitem{kondev3} J. L. Jacobsen and J. Kondev, Nucl. Phys. B \textbf{532},
635 (1998).

\bibitem{Pel} J. D. Pelletier, Phys. Rev. Lett. 78, 2672 (1997).

\bibitem{Barabasi}  A. L. Barabasi, R. Bourbonnais, M. Jensen, J.
Kertesz, T. Vicsek, and Y. C. Zhang, Phys. Rev. A \textbf{45}, R6951 (1992).

\bibitem{Herman}H. M. Srivastava, and J. J. Trujiilo, \textit{Theory and Applications of Fractional Differential Equations}, (Elsevier, Amsterdam, 2006).

\bibitem{kondev2}J. Kondev and C. L. Henley, Phys. Rev. Lett.
\textbf{74}, 4580 (1995).

\bibitem{Schramm} O. Schramm and S. Sheffield, Acta Mathematica, 202, 21 (2009), e-print arXiv:math.PR/0605337.

\bibitem{Schwartz} M. Schwartz redrived the KH result for the special
cases by Flory's argument: M. Schwartz, Phys. Rev. Lett. \textbf{86}, 1283 (2001).

\bibitem{Rajabpour} A. Saberi, M. A. Rajabpour, and S. Rouhani, Phys. Rev. Lett. \textbf{100},
044504 (2008); eprint arXiv:07122984.

\bibitem{saberi} A. A. Saberi, M. D. Niry, S. M. Fazeli, M. R. Rahimi Tabar, and S. Rouhani, Phys. Rev. E \textbf{77}, 051607 (2008).

\bibitem{Voss} R. F. Voss, in {\it Fundamental Algorithms for Computer Graphics},
edited by R. A. Earnshaw, {\it NATO Advanced Study Institute, Series
E: Applied Science} (Springer-Verlag, Heidelberg, 1985), Vol. 17,
p. 805; there are many methods to generate self-affine surfaces but
the accuracy of the above method was good enough for the range we
worked. One can find another method for broader range of roughness
exponents and other purposes in H. A. Makse, S. Havlin, M.
Schwartz, and H. E. Stanley, Phys. Rev. E \textbf{53}, 5445 (1996);
and H. Hamzehpour and M. Sahimi, Phys. Rev. E \textbf{73}, 056121 (2006).



\bibitem{movahed1} M. S. Movahed and S. Ghassemi, Phys. Rev. D {\bf 76}, 084037
(2007).

\bibitem{movahed2} F. Ghasemi, A. Bahraminasab, M. S. Movahed, S. Rahvar, K. R. Sreenivasan, and M. R. R. Tabar, J. Stat. Mech. P11008 (2006).

\bibitem{movahed3} M. S. Movahed and S. Rahvar, Phys. Rev. D {\bf 73}, 083518
(2006).

\bibitem{co04} Jr. R. Colistete, J. C. Fabris, S. V. B. Go\c{n}calves, and P. E. de
Souza, Int. J. Mod. Phys. D {\bf 13}, 669 (2004).
\bibitem{Cardy1} J. Cardy, \textit{Les Houches Summer School 1994} (North Holland,
Holland, 1996); eprint cond-mat/9409094.

\bibitem{Cardy2}  J. Cardy and R. M. Ziff, J. Stat. Phys. \textbf{110},
1 (2003).
\bibitem{Olami}Z. Olami and R. Zeitak, Phys. Rev. Lett. \textbf{76}. 247 (1996).

\bibitem{Mandelbrot} B. B. Mandelbrot, \textit{Fractals and Scaling in
Finance} (Springer, New York, 1997).

\bibitem{jan} N. Jan, D. Stauffer, and A. Aharony, J. Stat. Phys. \textit{92}, 325 (1998).

\end{thebibliography}
\end{document}